\documentclass[aps,reprint,superscriptaddress]{revtex4-1}
 \usepackage[utf8]{inputenc} 
 
 \usepackage{cmap}
\usepackage[T2A,T1]{fontenc}
\usepackage[utf8]{inputenc}
\usepackage[english]{babel}
\usepackage{textcomp} 
\usepackage{caption2} 
\usepackage{dsfont}
\usepackage{amssymb}
\usepackage[...]{youngtab}
\usepackage{bm}
\usepackage[unicode, pdftex]{hyperref}

\usepackage{amsmath,amsfonts,amssymb,amsthm,mathtools}

\newcommand{\bea}{\begin{eqnarray}}
\newcommand{\eea}{\end{eqnarray}}
\usepackage{tikz}
\usepackage{upgreek}
\usepackage{bbold}
\usepackage{amsmath}

\newcommand{\m}{\mu}
\newcommand{\n}{\nu}
\newcommand{\la}{\lambda}

\newcommand{\vf}{\varphi}
\newcommand{\al}{\alpha}
\newcommand{\be}{\beta}

\newcommand{\Ga}{\Gamma}

\newcommand{\om}{\omega}
\newcommand{\Om}{\Omega}

\newcommand{\de}{\delta}
\newcommand{\De}{\Delta}
\newcommand{\si}{\sigma}

\newcommand{\e}{\epsilon}
\newcommand{\ro}{\rho}

\newcommand{\et}{\eta}

\newcommand{\lb}{\left(}
\newcommand{\rb}{\right)}

\newcommand{\rar}{\rightarrow}

\def \beq{\begin{equation}}
\def \eq{\end{equation}}

\newcommand{\dd}{\partial}

\newcommand{\vna}{\bm \nabla}
\newcommand{\na}{\nabla}

\def\bmat{\begin{pmatrix}}
\def\emat{\end{pmatrix}}

\newcommand{\bear}{\begin{array}{c}}
\newcommand{\ear}{\end{array}}
\newcommand{\ds}{\displaystyle}
\newcommand{\bec}{\begin{cases}}
\newcommand{\ec}{\end{cases}}

\DeclareMathOperator{\rot}{rot}

\newcommand{\vx}{\bm x}

\newcommand{\vv}{\bm v}

\newcommand{\va}{\bm a}
\newcommand{\vA}{\bm A}
\newcommand{\vom}{\bm\Om}
\newcommand{\ve}{\bm E}
\newcommand{\vb}{\bm B}

\newcommand{\half}{\frac{1}{2}}

\newcommand{\sg}{\sqrt{-g}}
\newcommand{\og}{\frac{1}{\sqrt{-g}}}

\newcommand{\HH}{\mathcal{H}}

\begin{document}

\title{On chiral vortical effect in accelerated matter}

\author{P.G. Mitkin}
\affiliation{ITEP, NRC Kurchatov Institute, 117218 Moscow, Russia
}
\author{V.I. Zakharov}
\affiliation{ITEP, NRC Kurchatov Institute, 117218 Moscow, Russia
}

\affiliation{Pacific Quantum Center, Far Eastern Federal University, 690950 Vladivostok, Russia}

\begin{abstract}
We revisit the evaluation of the chiral vortical effect in the accelerated matter. 
To first order in the acceleration the corresponding matrix element of the axial
current can be reconstructed from the flat-space limit.
A crucial point is the existence in the this case of an extra conservation law
 of fluid helicity. As a result, the chirality of the microscopic degrees of freedom
and helicity of the macroscopic motion are separately conserved. This separation 
persists also in presence of gravity. Implications for the thermal chiral vortical effect
are discussed.
 
\end{abstract}

\pacs{}
\maketitle

\section{Introduction}

Chiral effects have attracted a lot of interest recently. 
These effects can influence the dynamics of a variety of systems, 
from interior of a neutron stars and quark-gluon plasma to Weyl and Dirac semimetals, 
for a review  see, e.g.,  \cite{Huang:2015oca,Kharzeev:2015znc}.
 Originally these effects were evaluated in the case of heated fermionic gas in presence of rotation and/or external magnetic field
%of the thermal 
%radiation from black holes and neutron stars 
\cite{Vilenkin:1979ui,Vilenkin:1980fu,Vilenkin:1980zv}. 
More recently, it was realized that in case of ideal fluid
 the chiral effects can be related to the quantum chiral anomaly  
\cite{Erdmenger:2008rm, Son:2009tf} 
and the corresponding transport coefficients
are explicitly calculable.

In absence of electromagnetic fields 
the axial current $j_5^\m$ receives a vortical contribution (chiral vortical effect, CVE):

\beq \label{hcur} j_5^\m = n_5 u^\m +  \si_\om \om^\m   \eq
where $n_5$ is the density of the microscopic constituents with axial charge, 
$u_{\mu}$
is the 4-velocity of an element of the fluid, 
and $\om^\m =  \half \e^{\m\n\al\be} u_\n \dd_\al u_\be$ 
is the vorticity vector. The vortical conductivity $\si_\om$ 
is directly related to the coefficient $C_{5}$ in front of the anomaly:

\beq \label{anomaly} \dd_\m j_5^\m = 
C_{5} \ve \cdot \vb \quad \rar \quad  \si_\om= 
\m^2 C_{5} + \frac{cT^2}{12}\eq
Here $\m$ is the chemical potential.
 For  a single
 right Weyl fermion $C_{5} = \frac{1}{4\pi^2}$ and $c=1$. 

The $T^2$ term in the r.h.s. of the Eq. (\ref{anomaly}) was obtained first
within the standard thermal field theory 
\cite{PhysRevD.20.1807}, but the discussion 
on its origin is still going on, see e.g. 
\cite{Landsteiner:2011cp,   
Golkar:2012kb, Avkhadiev:2017fxj,Glorioso:2017lcn, Flachi:2017vlp,
Landsteiner:2011iq,Stone:2018zel,Prokhorov:2020okl,Huang:2020kik,Alexandrov:2020zsj}. 
In particular, it is suggested to be related to the  gravitational anomaly:

\beq \label{genan}\na_\m j^\m_R = 
-\frac{1}{768\pi^2}\frac{\e^{\m\n\ro\si}}{\sqrt{-g}} R^\al_{\,\,\,\be\m\n}R^\be_{\,\,\,\al\ro\si} 
\eq
where $R_{\alpha\beta\gamma\delta}$ is the Riemann tensor. 
This relation is hard to check in the hydrodynamic approach. 
Indeed, the right-hand side of (\ref{genan}) is of the fourth order in gradients
and the corresponding current would be of the third order, 
while the  CVE is of the first order, see 
(\ref{hcur}). It was suggested, however, to interpret the 
generic temperature $T$ as the Unruh temperature $T_U=a/(2\pi)$ where $a$ 
is the gravitational acceleration (for details see \cite{Flachi:2017vlp,Stone:2018zel,Prokhorov:2020okl}).
 This substitution eliminates two of the gradients of the gravitational field 
in the hydrodynamic picture and  allows to link 
the $T^2$-part of CVE to the anomaly (\ref{genan}).

In addition to the  gravitational anomaly 
(\ref{genan}), there is another specific contribution to the divergence 
of the axial current in the external gravitational field 
(see \cite{Basar:2013qia} and references therein). 
Namely, it is argued that simultaneous account 
for the rotation and the acceleration results in the 
so-called gravimagnetic anomaly. 
The argumentation is based on the well-known analogy 
between rotation and the magnetic field. In more detail, 
one introduces the gravimagnetic fields 
$\ve_g = -\e\vna \vf$, $\vb_g = 2\e \vom$, where $\vf$ is the Newtonian potential, 
$-\vna \vf = \va_g$, and $\e$ is an energy of interacting test particle. 
Substituting this into (\ref{anomaly}) instead of the usual electromagnetic fields and summing over 
the Fermi sphere one gets, see \cite{Basar:2013qia} and references therein:

\beq  \label{newanomaly} \dd_\m j_5^\m =   
\frac{\m^2}{2\pi^2}\, (\va_g ,\, \vom ). \eq
Note that the r.h.s. of (\ref{newanomaly}) is of the second order in gradients.
Clarifying the physical meaning of (\ref{newanomaly}) is one of our aims here.

As is known, the chiral anomaly (\ref{anomaly}) 
 reveals instabilities in the chiral media with 
a nonzero axial chemical potential $\m_5 \neq 0$ 
\cite{Akamatsu:2013pjd, Khaidukov:2013sja, Kirilin:2013fqa, Avdoshkin:2014gpa, 
Manuel:2015zpa, Buividovich:2015jfa, Yamamoto:2015gzz, Hirono:2015rla, 
Kirilin:2017tdh, Li:2017jwv, Tuchin:2019gkg}. 
Namely, the state with $Q_5 \neq 0$ decays into helical magnetic fields. 
Generally speaking, both of the anomalies (\ref{genan}), (\ref{newanomaly}) can 
trigger similar phenomena. In particular, on the basis of (\ref{newanomaly}) 
one could argue that the state with a nonvanishing microscopic chirality $Q_5$ 
can be transformed into a helical macroscopic motion, 
even in absence of the electromagnetic interaction. If true, 
this would be a highly non-trivial theoretical phenomenon.

Studying the CVE in external gravitational field is 
also interesting for two other reasons. First, 
the phenomenology of heavy ion collisions indicates 
that the matter produced in these experiments undergoes 
accelerated expansion and this acceleration cannot be treated
 as a small gradient correction, see \cite{Abelev:2007zk} and further references in \cite{Kharzeev:2015znc}. Another point of interest in turning on gravitation 
is that it is known to mimic a temperature gradient \cite{Luttinger:1964zz}.

\beq \label{luttinger} \va_g \to -\frac{\vna T}{T}. \eq
The advantage of exploiting this relation is that the acceleration 
$\va_g$ can be introduced within a well-understood field-theoretic 
framework, while the notion of the temperature belongs to thermodynamics.

In this paper, we mostly focus on the gravimagnetic
 anomaly and its possible implications. 
We start with the review of the conservation 
laws of the ideal fluid in a flat spacetime and 
discussion of the nature of these laws in the section \ref{Sec:flat}. 
After that we turn to the hydrodynamics in the presence of a 
weak external gravitational field. We evaluate the terms linear 
in the gravitational acceleration following the standard procedure, 
see, e.g., \cite{Carroll:1997ar}. Namely, we start with the conservation laws 
in absence of the external gravitational field and then evaluate their 
modification by gravity. Following this procedure, we show that the
 Eq. (\ref{newanomaly}) is simply a manifestation of the equivalence
 principle and actually coincides with the Eq. (\ref{genan}) 
in absence of the $R\tilde{R}$ term. Throughout the text we use 
the metric signature $(-,+,+,+)$, and the normalization of the Levi-Civita 
symbol as $\e^{0123} = \sqrt{-g}$.

\section{Conservation laws for ideal fluids}
\label{Sec:flat}

Let us first consider ideal fluid in the absence of gravity. 
The hydrodynamics represents a universal 
approach based on the conservation laws, 
$\dd_\m j^\m_n = 0$, $\dd_\m T^{\m\n} = F^{\n\la} j_\la $, 
and  expansion in gradients.  

It is known, however, that in the case of ideal fluid there are 
extra conservation laws, not related to symmetries of the Lagrangian
 , see, e.g., 
\cite{Avdoshkin:2014gpa, 1987ApJ...319..207B}. 
The simplest example of such a law is  
conservation of the magnetic helicity $\HH_m$:

\beq \label{magnetic helicity} \bear \ds{\HH_m = 
\frac{1}{8\pi^2}\int d^3x \, \vA \cdot \vb } \\ \\ \ds{ \frac{d \HH_m}{dt} = 
-\frac{1}{4\pi^2}\int d^3x \, \ve \cdot \vb } \ear\eq 
If the chemical potential is a constant, 
 the electric field is screened inside an ideal conductor, and the magnetic helicity
 is conserved. Indeed, for an electric current 
$\bm J_{el} = \si_e \ve$ to be finite in the limit $\si_e \rar \infty$ the
 electric field inside the fluid should vanish. 

Let us emphasize 
that this conservation law is purely dynamical and is not related to a symmetry
of the Lagrangian. Rather, physics is sensitive to boundary conditions. 
This is a generic difference between the specific conservation
laws of the ideal fluid and Noether currents which are conserved locally.
Potentially, existence of the extra conservation laws is a source of an unusual
infrared sensitivity. We will come back to discuss this point later.

Chiral magnetic and chiral vortical effects are closely related to
the extra conservation laws. Indeed, with account for the quantum 
chiral anomaly the conserved axial charge for the system of right-handed 
fermions is given by

\beq \label{5} Q_R = Q^{naive}_R + \HH_m \eq

In the effective field theory describing the electromagnetic 
interaction of fermions in the hydrodynamic approximation 
\cite{Sadofyev:2010is}, the effective interaction of the microscopic degrees 
of freedom with photon field is shifted due to the chemical potential:

\beq \label{substitution0} A_\m \rar A_\m + \m u_\m \eq

If we substitute this into (\ref{5}), we get two additional 
hydrodynamic contributions to the axial charge, one being the 
fluid helicity $\HH_f =  \frac{1}{4\pi^2}\int d^3x\, \mu^2 \om^0 $, 
and the other is the  fluid-magnetic helicity $\HH_{mf} = 
\frac{1}{4\pi^2}\int d^3x\, \mu B^0$ with $B^\m = 
\half \e^{\m\n\al\be} u_\n F_{\al\be}$ 
being the magnetic field in the comoving frame 
of the element of fluid. Let us note that this substitution 
does not capture $T^2$-part of vortical effect. With account for that part, 
full axial charge is given by

\beq  \label{55} Q_R = Q^{naive}_R + \HH_m + \HH_f + \HH_{mf} + \HH_{tf} \eq
where thermal fluid helicity  $\HH_{tf}$ is the charge associated with the current

\beq j_{tf}^\m = \frac{1}{12}T^2 \om^\m \eq

Such an analysis that is based on consideration of hydrodynamic 
charges rather than spatial currents, albeit not new, is generally 
unnoted in the literature. However, it naturally leads to another observation 
that is worth to mention. The substitution (\ref{substitution0}) 
demonstrates the connection between the chiral vortical conductivity 
and the quantum anomaly existing 
even in  absence of the electromagnetic fields. 
However, it works only for the calculation of the 
contributions to the currents,  not their divergences. 
The point is that anomalous contributions to the currents 
are defined by the infrared properties of the system, while the divergences 
of anomalous currents are defined 
by the physics in the UV. Roughly speaking, if we try to reconstruct currents 
knowing their divergences the contributions discussed
appear as ``integration constants''.
In other words, the condition of vanishing of the divergences 
of the novel contributions to the currents becomes a constraint on
 the equation of state of the ideal fluid. 

Let us elaborate this point following \cite{Avdoshkin:2014gpa}.
First, consider the case of zero electromagnetic field, 
when the quantum anomaly is absent and the naive 
axial charge is in fact conserved. In that case only 
$\HH_f$ and $\HH_{tf}$ are non-zero in the equation (\ref{55}) 
apart from $Q_5^{naive}$. The divergence of the vorticity current is given by

\beq  \dd_\m \om^\m = \half \e^{\m\n\al\be} \dd_\m u_\n \dd_\al u_\be = 
\frac{1}{4} \om_{\al\be} \tilde{\om}^{\al\be}, \eq 
where $\om_{\al\be} =  \lb \dd_\al u_\be - \dd_\be u_\al \rb$ and 
$\tilde{\om}^{\al\be} = \half \e^{\al\be\m\n}\om_{\m\n} $ is the 
corresponding dual tensor. 
Introducing kinematic acceleration 
$a_\m = u^\ro \dd_\ro u_\m =  u^\ro \om_{\ro\mu}$ and 
algebraically decomposing $\om_{\m\n}$ in $a_\m$ and $\om_\m$ one finds: 

\beq \label{divrel} \dd_\m \om^\m = 2 a_\m \om^\m. \eq 
Note that the equation (\ref{divrel}) is purely kinematic. 
In order to account for the dynamics of the fluid we 
use the relativistic Euler equation:

\beq \label{Euler} a_\al = -\frac{1}{\e + p} P_\al^{\,\,\,\be} \dd_\be p ,\eq
where $\e$ is energy density, $p$ is pressure 
and the projector $P_{\al\be} = g_{\al\be} + u_\al u_\be$ 
satisfies $u^\al P_{\alpha\beta}=0$. 
Using that along with thermodynamical relations 
$dp = n d\m + s dT$ and $\e + p = n\m + sT$ (here $n$ is the 
density of the chosen kind of particles, right-handed in our case, 
and $s$ is the entropy density) one can rewrite (\ref{Euler}) as

\beq \label{afull} \bear \ds{a_\al =  - P_\al^{\,\,\,\be} \lb \frac{\dd_\be T}{T} + 
\frac{nT}{\e+p} \dd_\be \lb \frac{\m}{T} \rb \rb. }
\ear\eq 
It is convenient at that point to change the thermodynamic 
variables from $\m$, $T$ to $\frac{\m}{T}$ and $T$. 
In these variables the divergence of the fluid-helicity current is 

\beq \label{divfluid} \bear \ds{ 4\pi^2\dd_\al j^\al_{f} = 
\dd_\al \lb \m^2 \om^\al\rb } \\ \\ \ds{ = 2 \om^\al \lb \m T \dd_\al\lb\dfrac{\m}{T}\rb + \m^2\lb \dfrac{\dd_\al T}{T} + a_\al\rb \rb } \\ \\ \ds{ = \frac{2T^2\m s}{\e + p} \om^\al \dd_\al \lb \frac{\m}{T} \rb .  }\ear \eq
For this contribution to vanish at $T\neq0$ one needs
 to set $\om^\al \dd_\al \lb \frac{\m}{T} \rb$ zero. The condition
 $\dd_\al \lb \frac{\m}{T} \rb = 0$ is indeed satisfied 
in the thermal equilibrium in absence of external fields. 
Moreover, combining this condition with the equation (\ref{afull}) we see that 

\beq \om^\al a_\al = \om^\al \frac{\dd_\al T}{T}.\eq
As a result, the $T^2$ piece of the fluid helicity is also conserved:

\beq \dd_\al \lb T^2 \om^\al \rb = 2T^2 \om^\al \lb \frac{\dd_\al T}{T} 
- a_\al \rb = 0 .\eq
Since in  absence of electromagnetic fields 
the chiral anomaly actually vanishes, we deduce that the
$Q_5^{naive}$ is also conserved and there is no transition 
between these helicities in the dissipationless limit.

In case of non-zero electromagnetic interaction 
the picture modifies slightly. The condition $\si_{el} \rar \infty$ 
requires $\frac{E_\al}{T} - \dd_\al \dfrac{\m}{T} = 0$. 
Then for the various parts of the conserved hydrodynamic current we have:

\beq \label{divfluidE} \bear \ds{ 4\pi^2\dd_\al j^\al_{f} =  
2 \om^\al \lb \frac{T^2 s \m}{\e+p}\dd_\al\lb \frac{\m}{T}\rb + 
\frac{\m^2 nT}{\e+p} \frac{E_\al}{T} \rb } \\ \\ \ds{ = 2\m E_\al \om^\al ;  }
\ear \eq

\beq \bear  \ds{ 4\pi^2\dd_\al j^\al_{mf} =  
\lb \dd_\al \m - \m a_\al \rb B^\al - 2\m E_\al \om^\al } \\ \\ 
           
            \ds{ = E_\al B^\al - 2\m E_\al \om^\al.}
\ear \eq 

We see that the  sum of  the magnetic-, fluid- and mixed- helicities is conserved 
by itself:

\beq \label{combineddiv} \dd_\al j^\al_{m} + \dd_\al j^\al_{f} + \dd_\al j^\al_{mf} = 0, 
\eq 
which means that there is no production of the microscopic particles in the 
ideal-fluid regime and

\beq \dd_\al \lb n u^\al \rb = 0 ,\eq
even though the macroscopic helicities can interchange with each other. 
Let us note that this picture breaks with introduction of viscosity but 
as long as the ideal-fluid approximation is valid, it stays true. 

As for the $T^2$-contribution, it conserves by itself even in this case:

\beq \label{tf} \dd_\al \lb T^2 \om^\al \rb  = \frac{2nT^3}{\e+p} \lb \dd_\al \lb \frac{\m}{T}\rb - \frac{E_\al}{T} \rb \om^\al = 0 .\eq In particular, it seems that, in the ideal regime, there is no mechanism of transition between this part of fluid vorticity and the magnetic helicity, although vortical response in the photonic helical current is not zero \cite{Avkhadiev:2017fxj}.

Note that all of these additional conservation laws take 
place only in the stationary case. This brings up the question, 
what happens if we introduce some non-stationarity into the system. One way to do 
that is to consider the system in an external gravitational field, 
which is the focus of the next section. As was discussed in the Introduction, 
it seems that in that case there should be some additional inputs 
to the hydrodynamic charge. Another possible source of
 non-conservation is dissipative effects, which, in turn, can 
translate the non-stationarity caused by an external field into a stationary flow, 
which is characterised by corresponding transport coefficients. 
In principal, there should be some corrections to the hydrodynamic 
charges in higher orders in the gradient expansion as well.

\section{Turning on gravity}
\label{Sec:3}

What we would like to do now is to consider the case of a weak 
external gravitational field 
and study the effects linear in gravity. 
In particular, it would focus on  understanding the meaning of the 
gravimagnetic anomaly. 

The central point of our argument is that we can use 
the conservation laws in the flat space and rewrite them via equivalence principle
to include the efect of gravty. 
In particular,, in absence of electromagnetic fields we have for the fluid helicity:

\beq \label{divg} \frac{1}{4\pi^2}\na_\al \m^2\om^\al = 
\frac{1}{4\pi^2\sqrt{-g}}\dd_\al\lb \sg \m^2\om^\al \rb  = 0.  \eq
where $\nabla_{\alpha}$ is the covariant derivative.
Equation (\ref{divg}) imply that we work on a curved background. 
However, in the weak gravitational field approximation we can 
rewrite it as corrections to the equation $\dd_\m \om^\m = 0$ on the flat background. 
In more detail,  
we would like to consider an ideal fluid in the external weak gravitational field 
of a simple form with the only nonzero component

\beq \label{hgen} h_{00} = - 2b\eq
where the function $b = b (t,\,\vx)$ plays role of the gravitational potential. 
If in addition to the condition of the gravitational field to be weak we require 
the fluctuations of the velocity of the fluid to be small as well $u^i = v^i \ll 1$, 
then the 4-velocity of the fluid takes the form 
$u^\m = (1-b, \vv)$. Substituting this to the equation (\ref{divg}) we get:

\beq 
\ds{ \dd_\al j^\al_{f,(0)} = -\frac{\mu^2}{2\pi^2}\,\lb \vom \cdot \vna b\rb \,, } 
\eq 
with $\vom = \half \rot \vv$.

This approach can be generalized to the relativistic 
case as well. Indeed, writing the metric in the form 
(\ref{hgen}) says nothing about the state of the fluid 
in the geometry described by that metric. In fact, 
the choice (\ref{hgen}) actually fixes the frame we 
are working in without any reference to the fluid flow. 
In general, one has to specify the boundary conditions 
in order to proceed. In our setup, we would like to describe 
such fluid that each of its elements is accelerating. 
That means that the metric should have the form (\ref{hgen})
in the rest frame of each fluid element. In other words, we use the following ansatz:

\beq \label{h} h_{\m\n} = -2b\, u_\m u_\n, \eq 
with $b = -\half u^\m u^\n h_{\m\n}$. 

Let us stress that redefinition (\ref{h}), while looking harmless, 
is in fact a non-trivial procedure beyond the first order in gradients. 
Unlike the decomposition
 $F_{\m\n} = E_\m u_\n - E_\n u_\m + \e_{\m\n\al\be} u^\al B^\be$ 
which is identical and simply serves as definition of $E_\m$ and $B_\m$, 
in general case the equation (\ref{h}) genuinely makes $h_{\m\n}$ 
$u$-dependent and in that sense resembles the chemical 
shift (\ref{substitution0}). Much like the latter, which in 
case of uniform flow is simply a $U(1)$ gauge transformation, 
this decomposition makes the gravitational field gauge 
non-trivial as $u_\m$ becomes dynamic. %However, in contrast with $U(1)$ chemical shift, here it becomes important only in the second order by gradients simply because the Riemann curvature is of this order. One can say, that in the ideal regime the metric (\ref{h}) stays as pure gauge. 

We would also like to preserve some notion of this 
field to be stationary, or, to be more specific, ``comoving''-stationary. 
For that we require 

\beq \label{stat} u^\m \dd_\m b = 0. \eq 

Now we can proceed to expand the hydrodynamics variables in 
$h_{\m\n}$. We will denote the order in $h_{\m\n}$ by a subscript 
(or superscript) $(i)$. First let us note that change in geometry modifies 
the 4-velocity of the flow, which we can write as a series in $h_{\m\n}$: 
$u^\m = u^\m_{(0)} + u^\m_{(1)} + ...$. 
The function $u^\m_{(0)}$ would then play a role of a 
``flat'' fluid velocity. The corrections of the first order can be found as:

\beq \bear  \ds{ -1 = g_{\m\n} u^\m u^\n = 
\et_{\m\n}\lb u_{(0)}^\m u_{(0)}^\n + 2u^\m_{(0)} u^\n_{(1)}\rb - 2b} \\ \\
            \ds{ u^\n_{(1)} = -b u^\n_{(0)} .}
\ear\eq
Lowering the index we get

\beq \label{velocity_rescaling} u_\m = u^{(0)}_\m (1+b) +  \mathcal{O}(h^2)\,. \eq

Now we can use that and the equivalence principle to 
find the divergence of the zero-order currents. 
Start first with a naive current $nu^\m$. 
We saw that in the ideal regime in a flat 
space this current is conserved, therefore we expect that

\beq \label{usual current} 0 = \og\dd_\m\sg\, nu^\m  = 
\og \dd_\m nu^\m_{(0)} + \mathcal{O}(h^2), \eq
where we used the fact that $\sg = 1+b + \mathcal{O}(h^2)$. 
We see that the usual divergence of the naive current is zero.

We can proceed, actually, to see how the other equations 
of the ideal hydrodynamics modify in the external gravitational field. 
For the fluid acceleration we have

\beq \bear \label{afull1}
        \ds{a_\m = u^\n\na_\n u_\m = a^{(0)}_\m + u^{(0)}_\m u^\n_{(0)}\dd_\n b - u^{(0)}_\al u^\n_{(0)} \Ga^\al_{\n\m}} \\ \\
        \ds{ = a^{(0)}_\m - \half u^\n_{(0)}u^\ro_{(0)} \dd_\m h_{\n\ro} } \\ \\ 
        \ds{ = a^{(0)}_\m + \dd_\m b \,. } \\ \\ 
\ear \eq
Using that we can write the covariant form of the 
Euler equation (\ref{Euler}) in the form

\beq \label{newEuler} a_\m^{(0)} = -\frac{1}{\e+p}P_\m^{\,\,\,\n} \dd_\n p - \dd_\m b  .\eq
With that equation at hand we can repeat all the analysis of the preceeding section. 
For the divergences of the currents, however, it can be done using 
only kinematics, similar to the case equation (\ref{usual current}). 
For the fluid helicity we have:

\beq \label{fg} \bear 
        \ds{ \dd_\al j^\al_{f,(0)} = -\frac{\mu^2}{2\pi^2}\om^\al_{(0)} \dd_\al b\,. } 
\ear\eq
We see that this equation actually saturates 
the gravimagnetic anomaly (\ref{newanomaly}), 
and coincides with  the equation (\ref{genan}).
Note that we do not refer at all to the chiral gravitational anomaly.
but keep only terms 
with gravitational field coming from the Christoffel symbols 
and the definition of 4-velocity on a curved background. 

One can also demonstrate that this result is actually independent of 
the observer's rotation. To see this we can simply introduce the intrinsic 
rotation in the rest frame via the new parameter in the metric as 
$ h_{\m\n} = -2b\, u_\m u_\n - V_\m u_\n - V_\n u_\m$. 
Furthermore we fix the gauge freedom by requiring $V_\m u^\m = 0$, 
$u^\m \dd_\n V_\m = 0$. The latter condition ensures then that the equation 
(\ref{newEuler}) 
remains the same.  The Eq. (\ref{fg})  then looks like 

\beq \label{fgadd} \bear 
        \ds{ \dd_\al j^\al_{f,(0)}  = -\frac{\mu^2}{2\pi^2}\lb\om^\al_{(0)} - 
\Om^\al \rb \dd_\al b\,,}
\ear\eq
where $\Om^{\al} = \half \e^{\al\be\m\n}u^{(0)}_\be\dd_\m V_\n $ 
can be thought of as an intrinsic angular velocity of the observer's frame. 
This equation explicitly shows that the gravimagnetic anomaly, 
much like the CVE itself (see \cite{Liu:2018xip} for further discussion),
 is induced specifically by the fluid rotation and in this sense it is frame-independent. 
%This also brings attention to another issue:
% in general similar observation should be applicable to the acceleration
 %(see again equation (\ref{newEuler})), but since we use the equivalence principle 
%we already assume that the gravitational field in not screened by, 
%say, the temperature gradient in the fluid and the situation (from the ``flat'' point of view) is non-stationary.

Coming back to the case $V_\m = 0$,
 if there is an external electromagnetic field 
then from equations (\ref{combineddiv}) and (\ref{tf}) we have

\beq \bear\label{tfg} \ds{ \dd_\al j^\al_{tf,(0)}  = 
-\frac{T^2}{6}\om^\al_{(0)} \dd_\al b  } \ear \eq
\beq \bear \label{mfg}
            \ds{ \dd_\al j^\al_{f,(0)} + \dd_\al j^\al_{mf,(0)} + \dd_\al j^\al_{m,(0)} } \\ \\
            \ds{ = -\lb\frac{\mu}{4\pi^2}B^\al_{(0)} + 
\frac{\m^2}{2\pi^2} \om^\al_{(0)}\rb   \dd_\al b } 
            
\ear \eq
All these equations are a consequence of 
the equivalence principle and can be arrived at starting from the 
flat case through the substitution 

\beq \label{substitution} 
a_{kinematic} = a_{gravity},
\eq 
where $a_{gravity} = \dd_\al b $.  

Another use of the equivalence principle to recover the effects 
of the first order in the weak field expansion is the study of thermal effects. 
The idea is to use the analogy between the temperature 
gradient and the gravitational field. In other words, 
we could introduce the substitution

\beq \label{lutt2}
\dd_\m b = -a_\m^g = \frac{\dd_\m T}{T}. 
\eq 

Since the linear in $\dd_\m b$ terms are fixed by the equivalence principle, 
we expect that the terms linear in the $\frac{\dd_\m T}{T}$ should be fixed as well. 
This conjecture can be used to determine relation between different transport 
coefficients. We should note, however, 
that the substitution (\ref{lutt2}) should be used cautiously 
as it depends on the frame we are working in. 
In particular, the stationarity condition (\ref{stat}) 
implies that the 4-velocity satisfies 

\beq 
\frac{1}{T}u^\m \dd_\m T = 0
\eq 
which is a non-trivial condition. 

As an example of use of this substitution we turn to a 
non-dissipative effect reminiscent of a thermal Hall effect, 
recently discussed in \cite{Li:2020eon}. Namely, we discuss two 
second order contributions to the current of the form 

\beq \label{stephanov-current}
\de j^\m = c_1 \De^\m_{\,\,\,\ro} \dd_\n \om_\perp^{\n\ro}  - 
\frac{c_2}{T^2} \om_\perp^{\m\n} \dd_\n T,
\eq 
where $\De^\m_{\,\,\,\n} = \de^\m_\n + u^\m u_\n$, 
and $\om_\perp^{\m\n}$ is a part of $\om^{\m\n}$ that is orthogonal to $u^\m$: 

\beq 
\om_\perp^{\m\n} = \half\lb \dd^\m_\perp u^\n - \dd^\n_\perp u^\m \rb ,
\eq 
with $\dd^\m_\perp u^\n = \De^{\m\ro} \dd_\ro u^\n = \dd^\m u^\n + u^\m a^\n$, 
so that 

\beq 
\om_\perp^{\m\n} = \om^{\m\n} + \half u^\m a^\n - \half a^\m u^\n = - \e^{\m\n\al\be} u_\al \om_\be.
\eq 

In the simplest case of conformal theory dimensional analysis tells us that $c_1 = T f_1\lb \frac{\m}{T}\rb$ and $c_2 = T^2 f_2\lb \frac{\m}{T} \rb$. Since the second term in the equation (\ref{stephanov-current}) is
 linear in the temperature gradient, we expect that it could be recovered 
from the covariant form of the first term in (\ref{stephanov-current}) 
in the external gravitational field via the substitution (\ref{lutt2}). Indeed, 
using that $\om^\m = (1+b) \om^\m_{(0)} + \mathcal{O}(h^2)$, and 
$ \De^\m_{\,\,\,\ro} =  \De^\m_{\,\,\,\ro,(0)} + \mathcal{O}(h^2)$ , we write:

\beq \label{THE}\bear 
    \ds{ \De^\m_{\,\,\,\ro} \na_\n \om_\perp^{\n\ro} =  
\De^\m_{\,\,\,\ro} \og \dd_\n\lb  \sg\, \om_\perp^{\n\ro} \rb } \\ \\ 
    \ds{ =  (1+b)\De^\m_{\,\,\,\ro}\dd_\n \om_{\perp,(0)}^{\n\ro} - 
2 \om_{\perp,(0)}^{\m\n} \dd_\m b. }
\ear \eq 
Comparing equations (\ref{stephanov-current}) and (\ref{THE}) we can conclude that 

\beq 
c_2 = 2Tc_1
\eq 
This relation is different from the result obtained in \cite{Li:2020eon} 
for the specific case of conformal theory by the factor of $2$. This discrepancy may be related to the mentioned dependency 
of the substitution (\ref{lutt2}) on the hydrodynamic frame. 
Indeed, one can argue that, in case of the conformal theory, 
the effective description of the thermal gradients in terms of the 
gravitational fields should preserve the conformal symmetry, i.e. 
the metric (\ref{h}) should be replaced by the conformal metric 
$g_{\m\n} = (1+2b)\et_{\m\n}$. Note that this redefinition does
 not change the 4-velocity rescaling (\ref{velocity_rescaling}) as 
it is defined only by the projection of the metric on the velocity itself ($g_{00}$ 
in the comoving rest frame), and, therefore, 
the coefficients of the gravimagnetic anomaly stay the same. 
However, the expected contribution in (\ref{stephanov-current}) changes.
 Indeed, we now have $\om_\m = \om^{(0)}_\m + \mathcal{O}(b^2)$ and, 
repeating the derivation in (\ref{THE}), we have:
\beq\bear 
    \ds{ \De^\m_{\,\,\,\ro} \na_\n \om_\perp^{\n\ro} = 
 \De^\m_{\,\,\,\ro} \og \dd_\n\lb  \sg\, \om_\perp^{\n\ro} \rb } \\ \\ 
    \ds{ =  (1-3b)\De^\m_{\,\,\,\ro}\dd_\n \om_{\perp,(0)}^{\n\ro} -  
\om_{\perp,(0)}^{\m\n} \dd_\m b, }
\ear\eq 
which implies 
\beq 
c_2 = Tc_1,
\eq 
in agreement with \cite{Li:2020eon}. One possible explanation is that the effect of the centrifugal acceleration, which is compensated by the gradient of the temperature in the equilibrium, is not captured by the function $b$ when we use metric of the form (\ref{h}), meaning that the substitution (\ref{lutt2}) is valid only for the non-stationary gradients in that case.

 %We will comment on the possible limits of application of the  substitution (\ref{lutt2}) 
%some further in the conclusion.

%\beq \label{hgen} h_{\m\n} = \bmat -2b && \vV \\ \vV && 0 \emat \eq

%\beq \label{h} h_{\m\n} = -2b\, u_\m u_\n - V_\m u_\n - V_\n u_\m \eq 

%\beq \label{fg} \bear 
%        \ds{ \dd_\al j^\al_{f,(0)} = -\frac{\mu^2}{2\pi^2}\om^\al_{(0)} \dd_\al b - \frac{\m^2}{4\pi^2} \e^{\al\be\ro\si} \dd_\al V_\be \dd_\ro u^{(0)}_\si } \\ \\
%        \ds{ = -\frac{\mu^2}{2\pi^2}\om^\al_{(0)} \dd_\al b +  \frac{\m^2}{8\pi^2} a^{(0)}_\al \e^{\al\be\ro\si}u^{(0)}_\be \Om_{\ro\si} }
%\ear\eq

\section{Matching the Equivalence Principle and thermodynamics}
\label{Sec:4}
So far, while considering the fluid in the external gravitational field, we focused on the case when the gravity introduces non-stationarity into the picture. This, again, partially fixes the boundary conditions in our problem. Another approach is to consider the accelerating fluid in the equilibrium. To do that one must include acceleration into parameters characterizing matter in the equilibrium. The corresponding density operator $\hat{\rho}$ relevant to this case was worked out rather recently,
see \cite{Becattini:2017ljh}:
\begin{equation} \label{rho}
\hat{\rho}=\frac{1}{Z}\exp\Big(-\beta_{\alpha}\hat{P}^{\alpha}
-\frac{1}{2}\varpi_{\alpha\beta}\hat{J}^{\alpha\beta}\Big)
\end{equation}
where $\hat{P}^{\alpha}$ is the momentum operator, $\hat{J}^{\alpha\beta}$ are 
generators of the Lorentz rotations, $\beta_{\alpha}=\frac{u_{\alpha}}{T}$, and
$\varpi_{\alpha\beta}=\half\lb\partial_{\alpha}\beta_{\beta}-
\partial_{\beta}\beta_{\alpha}\rb$ is the thermal vorticity tensor. Note that, in contrast with $\om_{\al\be}$ which includes only the kinematic acceleration and vorticity of the fluid, $\varpi_{\al\be}$ also depends on the gradients of the temperature. Moreover, the Lorentz generators can be decomposed into the boost operators $\hat{K}$
and generators of  rotations, $\hat{J}$:
\begin{equation}
\hat{J}^{\alpha\beta}=u^{\alpha}\hat{K}^{\beta}-u^{\beta}\hat{K}^{\alpha}-
\epsilon^{\alpha\beta\gamma\delta}u_{\gamma}\hat{J}_{\delta}
\end{equation}
%Further details can be found in \cite{Becattini:2017ljh}.

Knowing the density operator $\hat{\rho}$ one can evaluate one-loop statistically
averaged matrix elements, such as matrix elements of the energy-momentum tensor
$T^{\alpha\beta}$ or of the axial current $J^{\alpha}_5$, see in particular
\cite{Prokhorov:2018qhq,Prokhorov:2019yft}. The novel point is that the boost operators are 
conserved but do not commute with the Hamiltonian \cite{Becattini:2017ljh}. If one does not introduce the 
vorticity, then it turns out that the matrix elements are polynomials in the acceleration $a$ and temperature $T$.
The whole calculation is in terms of the Minkowskian space, and the acceleration $a$ 
now is what we call kinematic acceleration, $a_{kinematic}$.

On the other hand, one can evaluate the same matrix elements associated with 
particles living on a non-trivial manifold with a boundary -- so-called conical manifold --
at the finite temperature. The acceleration is then introduced in geometrical terms and 
represents the gravitational acceleration.

It was demonstrated that, say, energy density $\rho_{s=1/2}$ 
for massless spin-1/2 particles 
is exactly the same within the both approaches, statistical and gravitational:
\begin{equation}
\rho_{s=1/2}=\frac{7\pi^2T^4}{60}+\frac{T^2a^2}{24}-\frac{17a^4}{960\pi^2}.
\end{equation}
This coincidence of the results can be considered as confirmation of the equivalence
principle. For further details see the original paper \cite{Prokhorov:2019yft}

\section{Discussion}

%To summarize, let us briefly recap the main points. In section \ref{Sec:flat} we review different hydrodynamic helicities that manifest themselves via chiral effects. Following \cite{Avdoshkin:2014gpa} we showed that on the equations of ideal fluid this helicities conserve by themselves or in some combination. The fact that the currents of these helicities are anomaly-induced leads to conjecture that even when we introduce non-stationarity, this currents should be conserved anyway, which can lead to a novel contributions to the chiral effects. That is in slight disagreement with a concept of gravimagnetic anomaly (\ref{newanomaly}) that has been appearing in the literature. 

In these notes we considered an application of the equivalence principle to
hydrodynamics of ideal fluid. From the very beginning, we put an emphasis on
the conservation laws inherent to the ideal fluid. The reason is that, indeed, the conservation
laws is a backbone of the hydrodynamic approach. Considering the fluid in the ideal regime allows to use the 
formalism of unitary field theories. Since we are interested in hydrodynamics we choose a kind
of effective field theory which describes physics of equilibrium. The corresponding 
effective vertices can readily be read off from the form of the density operator.

Simple manipulations within the effective field theory allow to re-establish a
fundamental for the ideal-fluid hydrodynamics observation that there are 
extra conservation laws which are highly non-local and not related to the symmetries of 
the Lagrangian. These are the conservation laws of various types of helical 
macroscopic motions.
In fact, these conservation laws have been discussed since very long in terms of 
classical theory. An unexpected turn of logic is that now they get related to 
quantum anomalies of the underlying field theories. While the anomaly is decided on
short distances, the helical currents emerge as an infrared completion of the theory.

As the next step, we utilize the equivalence principle to consider accelerated motion
of the fluid. The main point of our discussion is that it can be used to obtain effects linear in gravitational acceleration. A particular effect that can be recovered this way is so-called gravimagnetic anomaly. Formally, this result coincides with the result of ref. \cite{Basar:2013qia}. The interpretation is, however, 
different. From our point of view, the consequences from the equivalence principle
reduce to the classical observation $a_{kinematic}=a_{gravity}$ where $a_{kinematic}$
is associated with introduction of non-inertial coordinate system while $a_{gravity}$
is the acceleration induced by an external gravitational field.
The gravimagnetic anomaly appears not to be associated with any 
violation of the chiral symmetry. 

%We then proceed to study the flow of the fluid in the weak gravitational field in section \ref{Sec:3}. The main point of our discussion is the equivalence principle which can be used to obtain effects linear in gravitational acceleration. For that we use the metric of the form of (\ref{h}). The equivalence principle then can naively be understood as the substitution (\ref{substitution}). We should stress that this substitution defines the setup we are working in. The equivalence principle works straightforwardly only in the cases when the fluid is free-falling in the direction of the gravitational acceleration. If we look at equation (\ref{afull1}) we see that the equivalence principle simply states that the elements of the fluid follow the geodesics. That would not be the case if the gravitational field would be screened by, say, the temperature gradient, or if there would be some dissipation.

Although in our case \footnote{There exist examples \cite{Donoghue:1984ga,Buzzegoli:2021jeh} when the 
standard program of renormalization at finite temperature comes in contradiction with
the equivalence principle} application of the equivalence principle results 
in a routine answer, we encounter a kind of fundamental problem that the introduction
of the gravitational acceleration brings in non-stationarity. This particular issue becomes especially interesting if we go in another direction following \cite{Luttinger:1964zz}. If we introduce dissipative effects, this non-stationarity
 translates into a (stationary) transport phenomenon. 
The non-stationarity that external field brings in is important for this: say, 
in the stationary rotating fluid there should be a temperature gradient causing 
the centrifugal force but no transport along this gradient. 
This becomes especially apparent if we go to the reference 
frame rotating with a fluid, where the temperature gradient is 
screened by the gravitational force. In general it seems that the 
correct statement is that the dissipative transport is caused by the 
fields that are not screened in the ``comoving'' coordinate maps 
(see for example \cite{Dubovsky:2011sj}). This in turn leads us to a 
conjecture that a \textit{global} comoving coordinate map exists in 
general only for an ideal fluid and dissipative effects are in this sense 
similar to the introduction of genuine gravitational field. This also 
again indicates the specific nature of the conservation laws we are 
working with. For the ideal fluid, every fundamental conservation 
law should exist separately for each fluid element, and when the gravitational field, 
which itself should be invariant with respect to the coordinate diffeomorphisms, 
is applied, each volume element should behave independently. 
Conservation of the macroscopic fluid helicities, however, 
depends on the boundary conditions and does not fit into this picture. 
For now we leave this topic for a future discussion.

 Finally, in section \ref{Sec:4} we review recent developments in the study of the accelerating fluids in equilibrium. For that one can include the kinematic acceleration of the fluid into the density operator $\hat{\rho}$ (\ref{rho}) and use it to calculate statistically averaged matrix elements \cite{Becattini:2017ljh}, or do the same calculations on the curved background. The coincidence of the results obtained using these methods can serve as an additional confirmation of the equivalence principle. Our final remark is that within statistical approach linear in acceleration 
terms are absent. Indeed, these terms do not contribute to the matrix elements, see \cite{Becattini:2017ljh, Prokhorov:2019yft}. At first sight, this result
 is in variation with our fixation
of non-zero linear terms starting from the equivalence principle. In fact, there 
is no contradiction between the two approaches since they address different 
types of physics, equilibrium and non-equilibrium. Moreover, comparison
of these two types of calculations  
makes manifested dependence of the linear in acceleration terms on physics 
in far-infrared, which we emphasized in preceding sections.

\section{Acknowledgments}

The authors are thankful to O.V. Teryaev and A.V. Sadofyev 
for detailed discussions. The work on this paper has been partly
supported by RFBR grant No. 18-02-40056 and 
No. 0657-2020-0015 of the Ministry of Science and Higher 
Education of Russia. The work of P.M. 
is also supported by the Foundation for the Advancement of 
Theoretical Physics and Mathematics “BASIS” No. 20-1-5-134-1. \\

\bibliographystyle{bibstyle}
\bibliography{refs}

\end{document}